\documentclass[aps,prl,twocolumn,groupedaddress,floatfix]{revtex4-1}

\usepackage{epsfig}
\usepackage{subfigure}
\usepackage{graphicx}
\usepackage{amsfonts}
\usepackage[figuresright]{rotating}
\usepackage{amssymb}
\usepackage{amsmath}
\usepackage{psfrag}
\usepackage{esint} 
\usepackage{bm}
\usepackage[colorlinks,linkcolor=blue,anchorcolor=blue,citecolor=blue,urlcolor=blue]{hyperref}
\usepackage[version=4]{mhchem}

\def\be{\begin{equation}} \def\ee{\end{equation}}
\def\bea{\begin{eqnarray}} \def\eea{\end{eqnarray}}

\def\pp{\parallel}

\def\p{{\bf p}}

\renewcommand{\vec}[1]{\mathbf{#1}}

\newcommand{\up}{\uparrow}
\newcommand{\down}{\downarrow}

\newcommand{\acomm}[2]{\left\{#1,#2\right\}}
\def\bpm{\begin{pmatrix}} \def\epm{\end{pmatrix}}

\renewcommand{\vr}{\mathbf{r}}
\newcommand{\vk}{\mathbf{k}}

\newcommand{\vK}{\mathbf{K}}

\newcommand{\vq}{\mathbf{q}}
\newcommand{\vqp}{{\mathbf{q}^\prime}}

\newcommand{\FS}{\text{FS}}

\makeatletter
\newcommand*{\balancecolsandclearpage}{%
  \close@column@grid
  \clearpage
}
\makeatother

\begin{document}
\title{Vortices in a Monopole Superconducting Weyl Semi-metal}
\author{Canon Sun}
\affiliation{Department of Physics and Astronomy, The Johns
Hopkins University, Baltimore, Maryland 21218, USA}
\author{Shu-Ping Lee}
\affiliation{Department of Physics and Astronomy, The Johns
Hopkins University, Baltimore, Maryland 21218, USA}
\author{Yi Li}
\affiliation{Department of Physics and Astronomy, The Johns
Hopkins University, Baltimore, Maryland 21218, USA}
\date{\today}

\begin{abstract}
A monopole harmonic superconductor is a novel topological phase of matter with topologically protected gap nodes that result from the non-trivial Berry phase structure of Cooper pairs. In this work we propose to realize a monopole superconductor by the proximity effect between a time-reversal broken Weyl semi-metal and an $s$-wave superconductor. Furthermore, we study the zero-energy vortex bound states in this system by projection methods and by exact solutions. 
The zero modes exhibit a non-trivial phase winding in real space as a result of the non-trivial winding of the order parameter in momentum space. By mapping the Hamiltonian to the $(1+1)$d Dirac Hamiltonian, it is shown that the zero modes, analogous to the Jackiw-Rebbi mode, are protected by the index theorem.  Finally, we propose possible experimental realizations.
\end{abstract}
\maketitle

{\it Introduction.} -- The discovery of topological phases of matter has revolutionized our understanding of modern condensed matter physics. Unlike the traditional paradigm of Landau theory where states of matter are classified based on the symmetries they break, topological phases are characterized by topological invariants which are unchanged under continuous deformations \cite{Thouless1982,Haldane1988,Haldane1983,Avron1983,Niu1985,Kohmoto1985,Avron1990}. For example, the very first discovered topological phase, the integer quantum Hall state, has a non-trivial first Chern number, which arises due to the breaking of time-reversal (TR) symmetry \cite{Klitzing1980,Tsui1982}. The robustness of the Chern number is responsible for the quantized plateaus observed in measurements of Hall conductivity \cite{Laughlin1981,Thouless1983,Halperin1982}. Subsequently, the notion of a topological quantum number has been introduced to lattice systems, with band topology playing the central role in classifying novel quantum states of matter. This has led to the discovery of a wide range of topological materials, such as quantum anomalous Hall insulators \cite{Haldane1988,Yu2010,Chang2013,Liu2016}, topological insulators \cite{Kane2005a,Bernevig2006a,Bernevig2006,Fu2007a,Zhang2009,Xia2009,Chen2009,Hasan2010,Qi2011}, and Dirac and Weyl semi-metals \cite{Murakami2007,Wan2011,Burkov2011,Xu2011,Yang2011,Meng2012,
Witczak-Krempa2012,Hosur2012,Fang2012,Halasz2012,Cho2012,Son2013,Hosur2013,Wang2013a,Vazifeh2013,Nandkishore2014,Hosur2014,Potter2014,Haldane2014,Wei2014,Yang2014,Weng2015,Burkov2015,Borisenko2015,Xiong2015a,Xu2015,Xu2015a,Huang2015,Lv2015,Bednik2015,Lu2015,Li2018,
Xu2016,Yan2017,Armitage2018}.

One particular class of topological phases that has garnered significant attention recently is topological superconductors \cite{Read2000,Ivanov2001,Elliott2015,Kitaev2001,Stone2006,Tewari2007,Fu2008,Fu2009a,Chung2009,Lutchyn2010,Oreg2010,Cheng2010,Alicea2012,Elliott2015,Sato2017}. Perhaps the most striking feature of a topological superconductor is its ability to trap zero-energy Majorana modes in superconducting vortices. These Majorana zero-modes possess non-Abelian braiding statistics and are stable against small perturbations, which make them ideal candidates for realizing a topological quantum computer \cite{Ivanov2001,Stone2006,Nayak2008}. 
Furthermore, they can also be constructed from hybrid structures consisting of a topological insulator or semiconductor in proximity with an $s$-wave superconductor \cite{Fu2008,Linder2010,Fidkowski2011,Law2009,Qi2010b,Wimmer2010,Lutchyn2010,Qi2009a,Oreg2010}. The universal feature underlying all these systems is that their low-energy physics is described by a spinless chiral $p$-wave superconductor.

A recent work \cite{Li2018} has opened up the possibility for a superconducting gap function to possess pairing symmetry beyond the usual spherical harmonic symmetry. When the constituent electrons of a Cooper pair reside on two different Fermi surfaces (FSs) carrying opposite Chern numbers, the gap function possesses \emph{monopole harmonic symmetry} regardless of the concrete pairing mechanism. In this case the gap function is described by a monopole harmonic, as opposed to a spherical harmonic used in the traditional classification of unconventional superconductivity. Furthermore, the gap function has a non-zero total vorticity in momentum space over the FS. The non-zero vorticity forces the gap function to possess topologically protected nodes, in contrast to `traditional' topological superconductors, such as a two-dimensional $p_x+ip_y$ superconductor, which are fully gapped. This novel class of superconductors, called monopole superconductors, can be realized in, for example, a doped, TR-broken Weyl semi-metal in proximity with an $s$-wave superconductor, which has been studied in some previous works \cite{Li2018,Meng2012,Bednik2015,Lu2015,Cho2012}. This exotic pairing state is also closely related to the $J$-triplet pairing
of $p$-wave triplet pairing state with Cooper pair total angular momentum $J=1$, which has been proposed in systems of magnetic dipolar fermions \cite{Li2012b,Li2014a}.

In this work we study the zero-energy Majorana vortex bound states of a monopole superconductor. A simple model of a monopole superconductor in a TR-broken Weyl semi-metal is considered. Its gap function shows a non-trivial phase winding in momentum space. A string-like vortex in real space is then imposed onto the system and the wavefunctions of a branch of Majorana zero-energy vortex bound states are solved both analytically and numerically. 
The evolution of the zero mode wavefunction as the momentum along the vortex line varies is analyzed. 
It is shown by mapping the Hamiltonian to a $(1+1)d$ Dirac Hamiltonian with a mass domain wall the zero modes are protected topologically by the index theorem \cite{Tewari2007,Tewari2010,Jackiw1976}. 

\begin{figure}[htpb]
\subfigure[]{\centering  {\epsfig{file=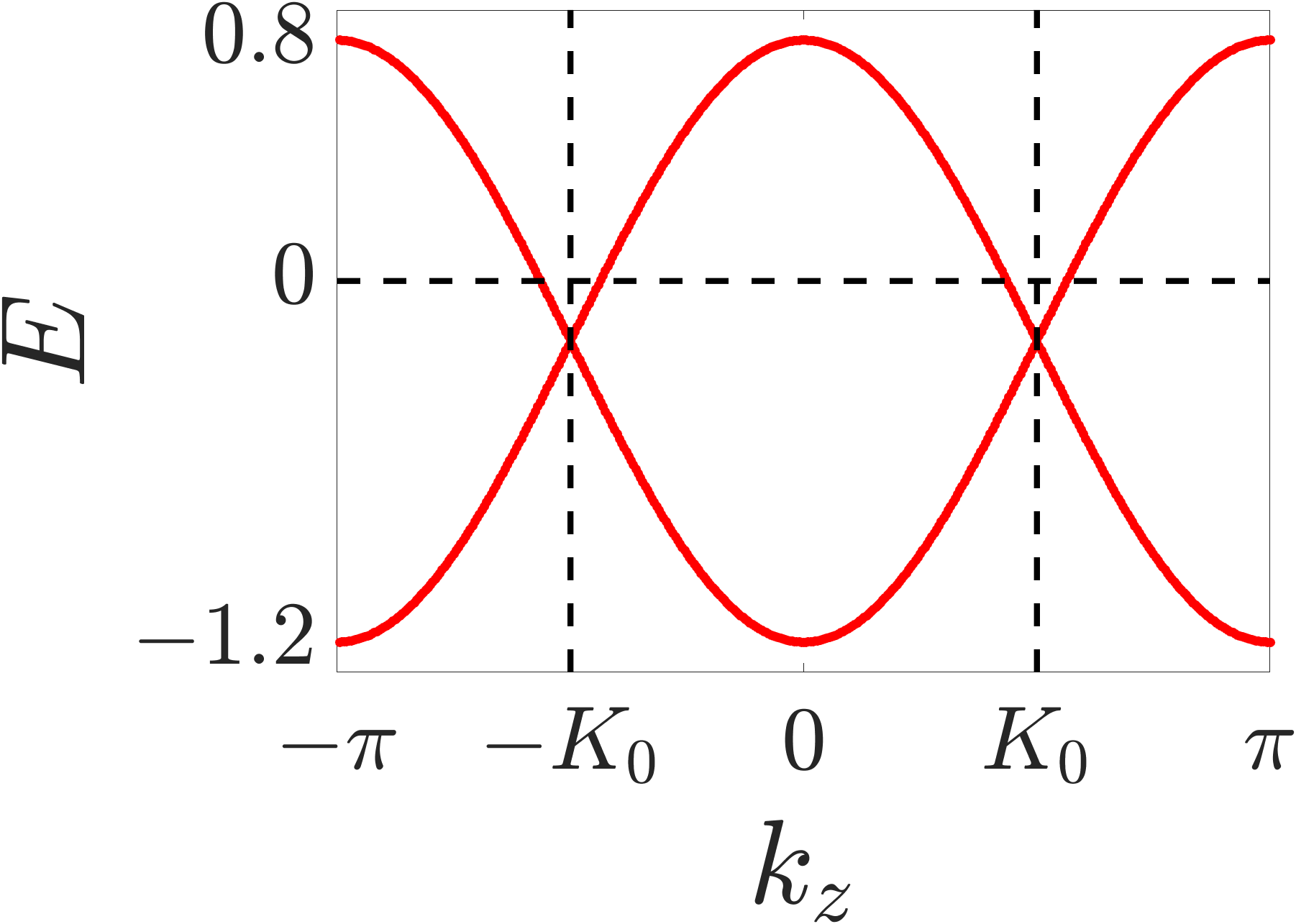,width=0.48\linewidth}}
}
\subfigure[]{\centering  {\epsfig{file=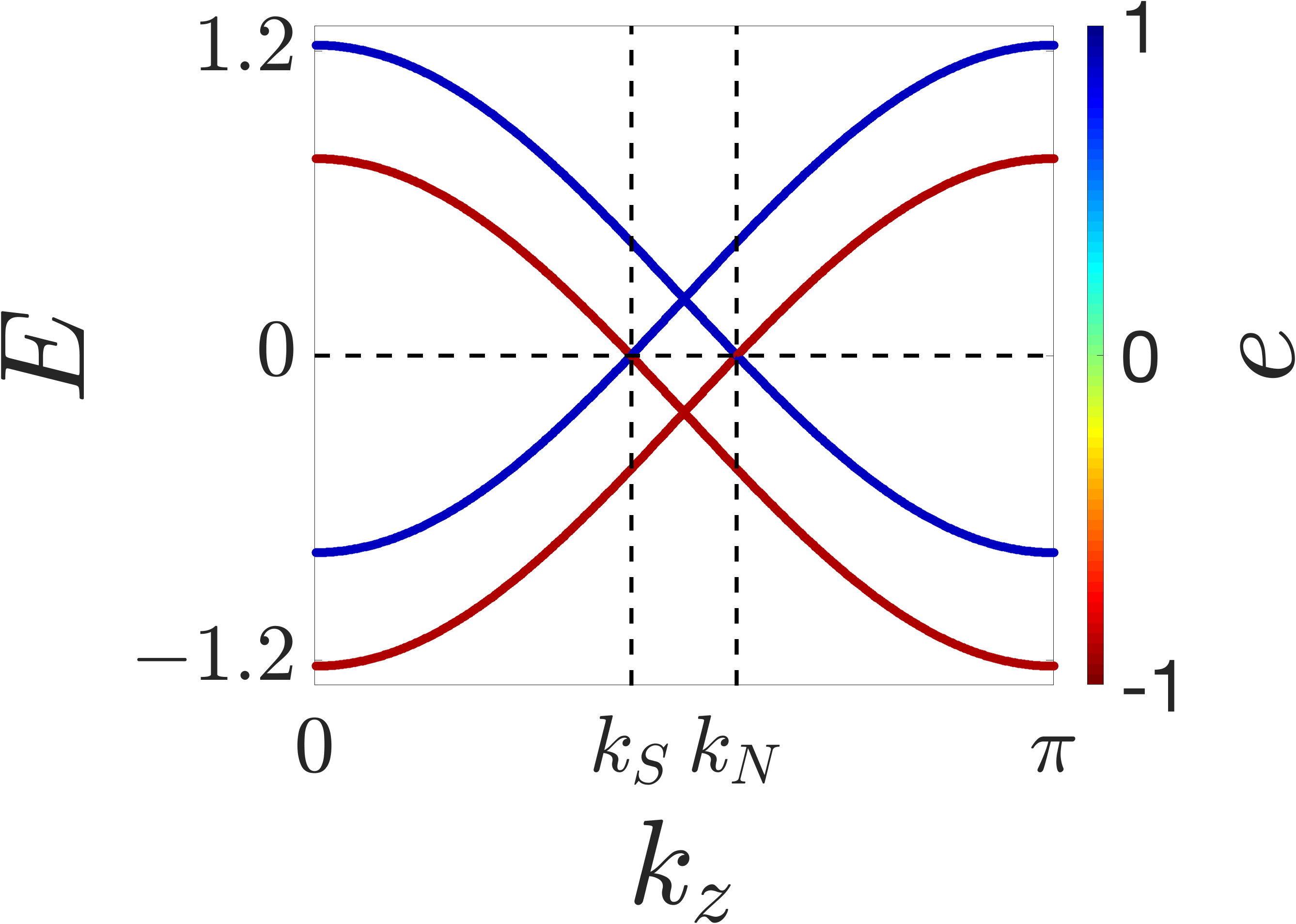,width=0.48\linewidth}}
}
\subfigure[]{\centering \epsfig{file=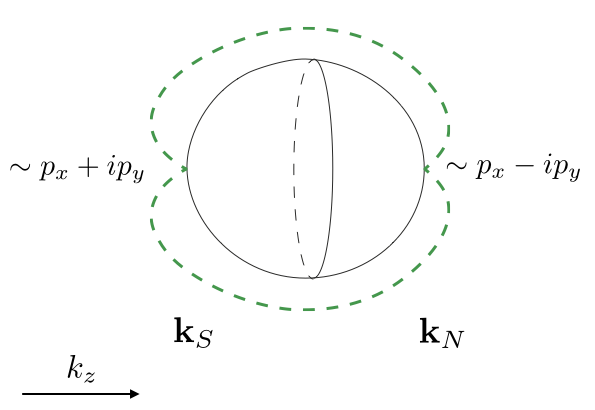,width=0.60\linewidth}}

\caption{($a$) Energy dispersion of the band Hamiltonian Eq. \eqref{eq:WeylHam}  along the $k_x=k_y=0$ cut showing the two Weyl nodes at $\pm \vK_0$. Parameter values are $t=1$ and $\mu=0.2$. ($b$)  Bulk BdG quasi-particle energy spectrum in the reduced Brillouin zone with $\Delta_0=0.1$ with the color representing the quasi-particle charge $e$.  ($c$) The magnitude of the
gap function (dashed green line) on the FS (solid black line) has nodes at the north and south poles, which manifest as the emergent Weyl nodes in ($b$). A momentum slice perpendicular to the $k_z$ direction near the north (south) pole is a 2D $p_x-ip_y$ ($p_x+ip_y$) superconductor. 
}
\label{fig:bandstructure}
\end{figure}

{\it Proximity-induced Monopole Superconductivity.} -- 
	A monopole superconductor can be realized in a doped TR-breaking Weyl semi-metal with proximity-induced superconductivity. For the sake of concreteness, let us consider the following model Hamiltonian, 
	\begin{equation}
    \label{eq:Ham}
    \hat{H}=\hat{H}_{\text{Weyl}}+\hat{H}_{\Delta},
    \end{equation}
    where $\hat{H}_{\text{Weyl}}$ is the Hamiltonian for the Weyl semi-metal and $\hat{H}_{\Delta}$ is the mean-field pairing Hamiltonian. The Weyl Hamiltonian reads 
	\begin{equation}
	\label{eq:WeylHam}
	    \hat{H}_{\text{Weyl}}=\sum_{\vk}\hat{c}^\dagger_{\vk}h(\vk)\hat{c}_{\vk},
	\end{equation}
	 where $\hat{c}_{\vk}=\left(\hat{c}_{\vk\up},\hat{c}_{\vk\down} \right)^T$ and $\hat{c}_{\vk\sigma}$ is the electron annihilation operator with spin $\sigma=\up$, $\down$. 
	 The matrix kernel $h(\vk)=\sum_{i=x,y,z}h_i(\vk) \sigma_i-\mu I$,
	 where
	 \begin{align}
	 \begin{split}
	     h_x(\vk)&=t \sin k_x,\ \ \, \ \ \, \ \ \, \ \ \, \ \ \, \ \ \, 
	     h_y(\vk)=t \sin k_y,\\
	     h_z(\vk)&=t\left(2-\cos k_x-\cos k_y-\cos k_z+\cos K_0\right).
	     \end{split}
     \label{eq:lattice}
	 \end{align}
    Here $\mu$ is the chemical potential, $\sigma_{x,y,z}$ are the Pauli spin matrices, $I$ is the identity matrix, and $t$ is the hopping amplitude. Without loss of generality, we assume $\mu>0$ throughout this article.
    This Hamiltonian originates from a three-dimensional tight-binding model on a cubic lattice with spin-orbit coupled nearest-neighbor hopping \cite{Yang2011,Cho2012,Lu2015}. It breaks TR symmetry but is invariant under the symmetry operation $(k_x,k_y,k_z)\mapsto (k_x,k_y,-k_z)$, and possesses two Weyl nodes along the $k_z$ axis at $\pm\vec{K}_0=(0,0,\pm K_0)^T$ with chiralities $\pm 1$. For simplicity, we restrict our discussion to the case with isotropic nodes by setting $t \equiv \hbar v_F$ and $K_0=\frac{\pi}{2}$. The energy dispersion of the model along the $k_x=k_y=0$ cut is shown in Fig. \ref{fig:bandstructure} ($a$). 
    The $s$-wave pairing Hamiltonian reads
    \begin{equation}
    \label{eq:pair}
    \hat{H}_{\Delta}=\sum_{\vk}{\vphantom{\sum}}'\hat{c}^\dagger_{\vk}\Delta_0 i\sigma_y \hat{c}^\dagger_{-\vk}+\text{H.c.},
    \end{equation}
    where the sum over $k_z$ is only for $k_z>0$, as indicated by the prime symbol in the sum. Here the $s$-wave pairing amplitude $\Delta_0$ is taken to be real. 
    
    The Bogoliubov-de Gennes (BdG) quasi-particle spectrum along the $k_x=k_y=0$ cut is plotted in Fig. \ref{fig:bandstructure} (b), which shows two nodes at $k_z$ equal to $k_N=K_0+ q_{\text{node}}$ and $k_S=K_0-q_{\text{node}}$, where $q_{\text{node}}=\sqrt{q_F^2+(\Delta_0/\hbar v_F)^2}$ and the Fermi wavevector $q_F\equiv \mu/(\hbar v_F)$. Further details of the BdG Hamiltonian and quasi-particle excitations are presented in the Supplemental Material (S. M.) I.


	{\it Low-energy Effective Hamiltonian.} --  To illustrate the monopole harmonic pairing, 
	we construct a low-energy effective theory for the Hamiltonian Eq. \eqref{eq:Ham}. The doping $\mu>0$ creates two Fermi surfaces $\FS_{\pm}$ carrying Chern numbers $\pm 1$ about the Weyl nodes at $\mp \vec{K}_0$. The pairing \emph{between} $\FS_+$ and $\FS_-$ gives the Cooper pairs a non-trivial Berry phase structure and is characterized by a monopole harmonic function. 
	 With small doping, the low-energy Hamiltonians around $\pm \vec{K}_0$ are represented by 
		\begin{equation}\label{eq:WeylnodeHam}
		h_{\pm }(\vq)\equiv h(\pm \vec{K}_0+\vq)\approx \hbar v_F( q_x\sigma_x+q_y\sigma_y\pm q_z\sigma_z)-\mu I,
		\end{equation}
		where $\vq$ is the wavevector relative to the respective nodes. 
		The electrons in the inter-FS Cooper pair are individually described by the eigenstates $\xi_-(\vq)=(u_{\vq},v_{\vq})^T$ and $\xi_+(-\vq)=(u_{\vq}, -v_{\vq})^T$, where $u_{\vq}=\cos(\theta_{\vq}/2)$ and $v_{\vq}=\sin(\theta_{\vq}/2) e^{i\phi_{\vq}}$. This choice of gauge,  which we term gauge I, has the Dirac string along $\theta_{\vq}=\pi$. The single-electron creation operators on the helical Fermi surfaces $\FS_{\mp}$ are defined as $\hat{\alpha}_{\mp}^\dagger(\pm \vq)=\sum_{\sigma}\xi_{\mp,\sigma}(\pm\vq)\hat{c}^\dagger_{\pm \vec{K}_0\pm\vq,\sigma}$.
		
	Even though Eq. \eqref{eq:pair} describes simple on-site, singlet, $s$-wave pairing, the Cooper pairing wavefunctions acquire a non-trivial Berry phase structure and the projected gap function is characterized by a monopole harmonic
function. Consider an inter-FS pair state consisting of two electrons  with wavevectors $\vk=\pm (\vK_0+\vq)$ living on $\FS_{\mp}$. 
		In the weak-coupling regime $|\Delta_0|\ll |\mu|$, the pairing Hamiltonian can be projected onto the helical FSs, yielding
	\begin{equation}
	    \hat{\tilde{H}}_{\Delta}=\sum_{\vq}{\vphantom{\sum}}'\hat{\alpha}^{\dagger}_-(\vq)\tilde{\Delta}(\vq)\hat{\alpha}^\dagger_{+}(-\vq)+\text{H.c.},
	\end{equation}
 where the projected gap function is $\tilde{\Delta}(\vq)=-\Delta_0 2u_{\vq}^*v_{\vq}^*=-\Delta_0\sin\theta_{\vq}e^{-i\phi_{\vq}}$, exhibiting nodes at the north and south poles. This expression is valid everywhere except at the south pole where the Dirac string lies. In order to include the south pole in the description, one must employ a different gauge. For example, one could choose the following gauge, which we term gauge II: $u_{\vq}=\cos(\theta_{\vq}/2)e^{-i\phi_{\vq}}$ and $v_{\vq}=\sin(\theta_{\vq}/2)$. In gauge II, the Dirac string is along $\theta_\vq=0$ and the projected gap function is $\tilde{\Delta}(\vq)=-\Delta_0\sin\theta_{\vq}e^{i\phi_{\vq}}$. The fact that the gap function must be described using two different gauge patches is one of the defining characteristics of a monopole superconductor. The gap function can be written more compactly by employing the monopole harmonic functions $Y_{qlm}(\theta_\vq,\phi_\vq)$ \cite{Wu1976,Haldane1983,Li2018}, in terms of which the gap function in our model is $\tilde{\Delta}(\vq)=-\Delta_0 \sqrt{\frac{8\pi}{3}}
Y_{-1,1,0}(\theta_\vq,\phi_\vq)$.

After projection, the Hamiltonian reads $\hat{\tilde{H}}=\sum_{\vq}^\prime\hat{\tilde{\Psi}}^\dagger(\vq)\tilde{H}_{\text{BdG}}(\vq)\hat{\tilde{\Psi}}(\vq)$, where we have defined the two-component spinor $\hat{\tilde{\Psi}}(\vq)=\left[\hat{\alpha}_-(\vq),\hat{\alpha}^\dagger_+(-\vq) \right]^T$ and the one-body BdG Hamiltonian is
\begin{equation}\label{eq:BdG Ham}
    \tilde{H}_{\text{BdG}}(\vq)=\begin{bmatrix}
    -\mu +\hbar v_F |q|&\tilde{\Delta}(\vq)\\
    \tilde{\Delta}^*(\vq)&\mu-\hbar v_F|q|
    \end{bmatrix}.
\end{equation}
Consider a momentum space cut at a fixed $k_z$ near the north pole. Under gauge I, which is regular at the north pole, the gap function is $\tilde{\Delta}(\vec{q})=-\Delta_0\frac{1}{|q|}(q_x-iq_y)$. In this case, $\tilde{H}_{\text{BdG}}(\vq)$ is equivalent to a two-dimensional (2D) spinless $p_x-ip_y$ superconductor. A chiral $p$-wave superconductor has two distinct phases: the topological non-trivial weak-pairing and topologically trivial strong-pairing phases. These two situations correspond to whether the momentum space cut of $k_z$ meets FS$_+$, or, not. For the cut near the south pole, we employ gauge II, giving $\tilde{\Delta}(\vq)=-\Delta_0\frac{1}{|q|} (q_x+iq_y)$, which is equivalent to a 2D spinless $p_x+ip_y$ superconductor.
A schematic diagram of $\tilde{\Delta}(\vq)$ is shown in Fig. 
\ref{fig:bandstructure} ($c$).
	
	{\it Majorana Vortex Bound States.} -- We now consider an inhomogeneous monopole superconductor with a vortex line situated at $\vr_\parallel=(x,y)^T=(0,0)^T$. Because translational symmetry in the $z$ direction is preserved even in the presence of a vortex, $k_z$ remains a good quantum number, and the pairing Hamiltonian can be written as
\begin{equation}
    \hat{H}_{\Delta}=\sum_{k_z}\int d^2r_{\parallel} \hat{\psi}^
\dagger_{k_z}(\vr_\parallel)\Delta(\vr_\parallel)i\sigma_y \hat{\psi}^
\dagger_{-k_z}(\vr_\parallel)+\text{H.c.},
\end{equation}
where $\hat{\psi}_{k_z}(\vr_\parallel)=\left[\hat{\psi}_{k_z,\up}(\vr_\parallel),\hat{\psi}_{k_z,\down}(\vr_\parallel) \right]^T$ and we have taken the continuum limit on the $x$-$y$ plane. A vortex configuration with winding number $+1$ is imposed on the $x$-$y$ plane: $\Delta(\vr_\parallel)=\Delta(r_\parallel) e^{i\phi_\vr}$, where $r_\parallel=\sqrt{x^2+y^2}$ and $\Delta(r_\parallel)=\Delta_0\tanh \left(r_\parallel/\xi\right) $ is set to describe the radial profile of the gap function and $\xi$ is the healing length. The detailed vortex solution needs to be solved self-consistently. 
Nevertheless, the topological properties, such as the existence of the zero-energy vortex bound states, should only depend on the phase winding of $\Delta(\vr_\parallel)$ and is independent of the self-consistency procedure.

We seek vortex bound state solutions to the Hamiltonian Eq. \eqref{eq:Ham}. In the weak coupling limit $\Delta_0\ll \mu$, the coherence length $\xi=\hbar v_F/\Delta_0$ is much smaller than the in-plane Fermi wavelength $\lambda_{F,\parallel} =2\pi/k_{F,\parallel}$, where $k_{F,\parallel}=\mu\sin\theta_\vq$, so long as we are away from the north and south poles. Hence, for vortex excitations we can approximate the gap function as $\Delta(r_\parallel)\approx (\Delta_0/\xi)r_\parallel$. Under this approximation, the pairing Hamiltonian can be Fourier transformed to momentum space as
\begin{equation}
    \hat{H}_{\Delta}=\sum_{k_z>0}\int \frac{d^2q_\parallel}{(2\pi)^2} \hat{c}^\dagger_{k_z}(\vq_{\parallel}) \Delta(\vq_\parallel)i\sigma_y\hat{c}^\dagger_{-k_z}(-\vq_\parallel)+\text{H.c.},
\end{equation}
where $\vq_\parallel=(q_x,q_y)^T$ is the in-plane wavevector,  $q_{\parallel}=\sqrt{q_x^2+q_y^2}$, $\Delta(\vq_\parallel)=\frac{\Delta_0}{\xi} Z(\vq_\parallel)$, $Z(\vq_\parallel)=ie^{i\phi_\vq}\left(\partial_{q_\parallel}+\frac{i}{q_\parallel}\partial_{\phi_\vq} \right)$ is the wavevector representation of the position operator $\hat{x}+i\hat{y}$. As before, we project the pairing Hamiltonian onto the FSs, which yields $\hat{\tilde{H}}_{\Delta}=\sum_{q_z}^\prime\int \frac{d^2q_\parallel}{(2\pi)^2} \hat{\alpha}^\dagger_-(\vq)\tilde{\Delta}(\vq)\hat{\alpha}^\dagger_+(-\vq)+\text{H.c.}$, where the projected gap function is
\begin{equation}\label{eq:projected Ham}
    \tilde{\Delta}(\vq)=\frac{\Delta_0}{\xi}\left(\tilde{Z}(\vq)+\tilde{A}(\vq)\right),
\end{equation}
the projected position operator is $\tilde{Z}(\vq)=-2u^*_\vq v^*_\vq Z(\vq)$, and $\tilde{A}(\vq)=-u_{\vq}^*Z(\vq)v^*_\vq-v^*_{\vq}Z(\vq) u_{\vq}^*$.

The BdG Hamiltonian takes the form given in Eq. \eqref{eq:BdG Ham}, but with the gap function replaced with Eq. \eqref{eq:projected Ham}. Since only low-energy excitations are of interest, we expand the in-plane wavevector about the FS by writing $q_\parallel=k_{F,\parallel}+\delta q_\parallel$, we obtain
\begin{equation}
    \tilde{H}_{\text{BdG}}(\vq)=\begin{bmatrix}
    \hbar \tilde{v}_F \delta q_{\parallel}&\tilde{\Delta}(\vq)\\
    \tilde{\Delta}(\vq)&-\hbar \tilde{v}_F\delta q_\parallel
    \end{bmatrix},
\end{equation}
where $\tilde{v}_F =v_F\sin\theta_\vq$ is the in-plane Fermi velocity. In gauge I, $\tilde{Z}(\vq)\approx -i\sin\theta_\vq\left(\partial_{\delta q_{\parallel}}+\frac{i}{k_{F,\parallel}}\partial_{\phi_\vq}  \right)$ and $\tilde{A}(\vq)\approx -i\frac{\sin\theta_\vq}{2k_{F,\parallel}}\left(1+\cos^2\theta_\vq \right)$. We seek zero-energy solutions to the BdG Hamiltonian. To that end, it is beneficial to discuss the symmetries possessed by the projected BdG Hamiltonian. First, it satisfies $\tilde{H}_{\text{BdG}}(\vq_\parallel,k_z)=\tilde{H}_{\text{BdG}}(\vq_\parallel,-k_z)$, which is inherited from the band structure Hamiltonian. Secondly, it possesses, by construction, charge conjugation symmetry $\mathcal{C}\tilde{H}_{\text{BdG}}(\vq_\parallel,k_z)\mathcal{C}^{-1}=-\tilde{H}_{\text{BdG}}(-\vq_\parallel,-k_z)$, where $\mathcal{C}=\tau_x K$ and $K$ is the complex conjugation operator. Furthermore, it satisfies $\tilde{H}_{\text{BdG}}(\vq_\parallel,k_z)=\tilde{H}_{\text{BdG}}(-\vq_\parallel,k_z)$, which arises as a result of the mixing between the phase windings in real and momentum space.
Therefore, for every solution $\psi_{k_z}(\vq_\parallel)$ with energy $E_{\vq_\parallel,k_z}$, there exists another solution, $\mathcal{C}\psi_{k_z}(\vq_\parallel)$ with energy $-E_{\vq_\parallel,k_z}$. The zero mode solutions can be arranged to satisfy $\psi_{0,k_z}(\vq_\parallel)=\mathcal{C}\psi_{0,k_z}(\vq_\parallel)$. In other words, we look for a zero mode solution of the form $\psi_{0,k_z}(\vq_\parallel)=[f_{k_z}(\vq_\parallel),f^*_{k_z}(\vq_\parallel)]^T$. In gauge I, we find the solution to be
\begin{equation}
    \psi_{0,k_z}(\vq_{\parallel})=e^{-\frac{b^2}{2}\left(\delta{q_\parallel}+\frac{c}{b^2}\right)^2}\begin{bmatrix}
    e^{-i\frac{\pi}{4}}\\
    e^{i\frac{\pi}{4}}
    \end{bmatrix}
\end{equation}
where $b^2=\hbar v_F\xi/\Delta_0$ and $c=(1+\cos^2\theta_\vq)/(2k_{F,\parallel})$. Under our approximation, $b\gg c$. Reverting to the spin basis and taking the Fourier transform on the $x$-$y$ plane, the zero mode wavefunction is found to be, for $\lambda_{F,\parallel}\ll r_{\parallel}\ll \xi$, 
\begin{equation}\label{eq:zero-energysolutiondecay}
\psi_{0,k_z}(\vr_\parallel)\approx e^{-\frac{r_\parallel^2}{2b^2}}\chi_{k_z}(\vr_\parallel),
\end{equation}
where $\psi_{k_z}(\vr_\parallel)=[u_{0,k_z\up},u_{0,k_z\down},u^*_{0,k_z\up},u_{0,k_z\down}^*]^T$ and $\chi_{k_z}(\vr_\parallel)$ are four component spinors in the (particle-hole)$\otimes$(spin) basis. The spinor $\chi_{k_z}(\vr_\parallel)$ is 
\begin{equation}\label{eq:zero-energysolution}
\chi_{k_z}(\vr_\parallel)=
\begin{bmatrix}
e^{-i\frac{\pi}{4}}\cos\frac{\theta_\vq}{2} J_0(k_{F,\parallel}r_\parallel)\\
e^{i\frac{\pi}{4}}\sin\frac{\theta_\vq}{2} e^{i\phi_\vr}J_1(k_{F,\parallel}r_\parallel)\\
e^{i\frac{\pi}{4}}\cos\frac{\theta_\vq}{2} J_0(k_{F,\parallel}r_\parallel)\\
e^{-i\frac{\pi}{4}}\sin\frac{\theta_\vq}{2} e^{-i\phi_\vr}J_1(k_{F,\parallel}r_\parallel)
\end{bmatrix}.
\end{equation}
where $J_{n}(z)$ is the Bessel function of order $n$. This solution can be understood in the following way. When $|q_z|<q_F$, each momentum slice with a fixed $q_z$ cuts out a Fermi surface cross section, which is effectively a two-dimensional topological superconductor with a full gap, hence it can host a single zero-energy Majorana vortex bound state \cite{Read2000}. For $k_z$ greater than $k_N$ or less than $k_S$, i.e. $|q_z|>q_{F}$, the momentum slice does not cut the Fermi surface anymore.  We enter the topologically trivial strong-pairing phase which does not have zero-energy solutions. 

The zero mode solution can be also be obtained without Fermi surface projection, as detailed in S.M. II. The solution in S.M. II describes all the zero modes even after Fermi surface reconstruction, whereas the present analysis only captures zero modes inside the original FS. More specifically, when $q_F<|q_z|<q_{\text{node}}$, there are non-oscillatory solutions, as given in Eqs. \eqref{eq: exact exponential} and \eqref{eq: exact I}.

The phase windings and the relative weights of different components of Eq. \eqref{eq:zero-energysolution} can be understood from the projected
gap function $\tilde{\Delta}(\vq)$. Consider a momentum slice near the north pole. The corresponding helical states at $\mu>0$ are approximately
spin polarized along the $z$-direction. The gap function vorticity around the north pole is approximately equivalent to a two-dimensional $p_x-ip_y$ superconductor. Hence, the 1st and 3rd rows of Eq. \eqref{eq:zero-energysolution} dominate. A similar problem has been solved in the previous works, such as Ref. \cite{Cheng2010}, and the particle part of the wavefunction exhibits no phase winding. A similar argument applies near the south pole: The helical states are nearly spin down, and the pairing is approximately equivalent to a two-dimensional $p_x+ip_y$ superconductor. Hence, the 2nd and 4th rows of Eq. \eqref{eq:zero-energysolution}  dominate and exhibit the phase windings $e^{\pm i\phi_\vr}$. The solutions at the north and south poles are not related by symmetry: The momentum space gap function vorticities are opposite at the north and south poles, but the real space gap function winding number remains the same. As $k_z$ is varied continuously from the north pole to the south, since no gap function nodes are swept, the phase winding of each component of of Eq. \eqref{eq:zero-energysolution} does not change, but the dominant components change from 1 and 3 to 2 and 4.

\begin{figure}[tbp]
\subfigure[]{\centering \epsfig{file=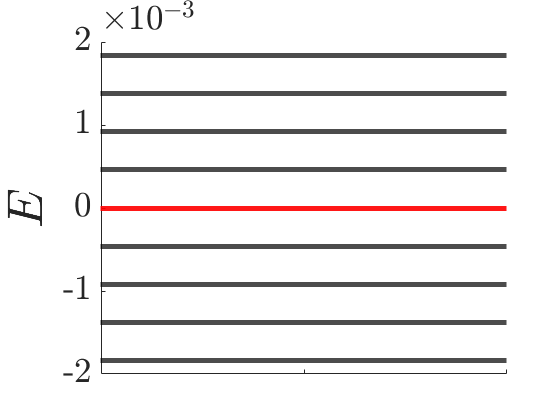,width=0.32\linewidth}}
\subfigure[]{\centering\epsfig{file=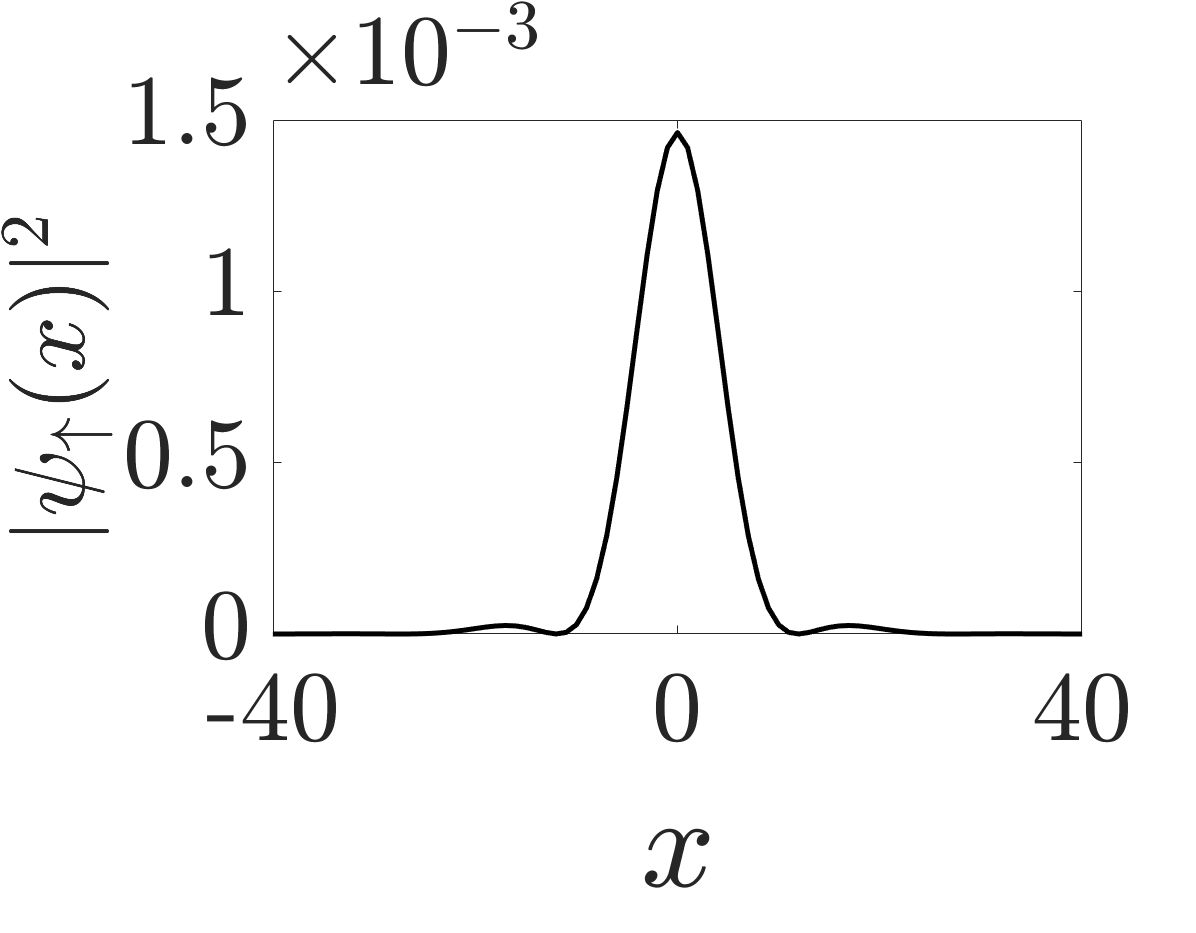,width=0.32\linewidth}}
\subfigure[]{\centering\epsfig{file=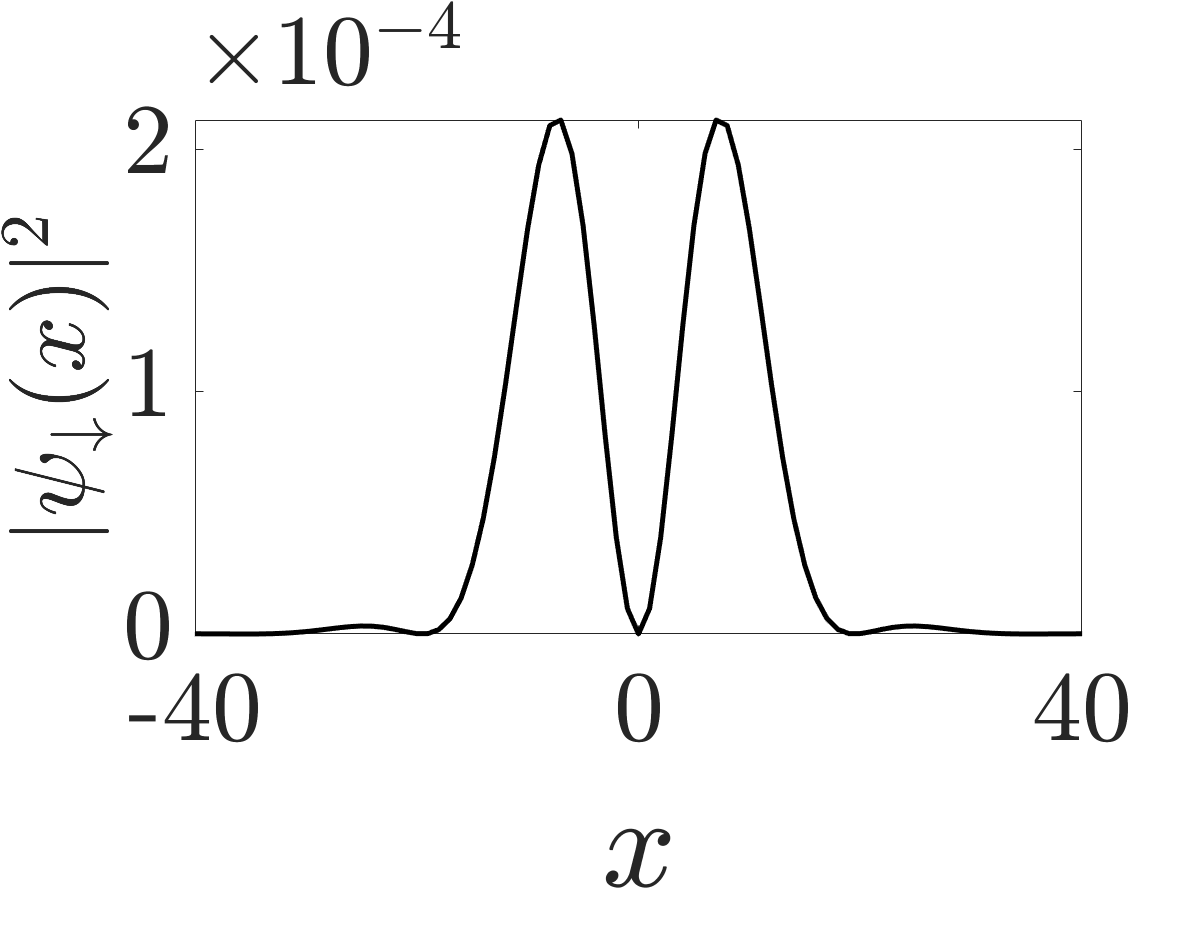,width=0.32\linewidth}}
\caption{
($a$) The energy spectra of the lowest lying energy states with a vortex for the cut of $q_z=0$, showing a zero-energy mode. The radial wavefunctions of the zero mode ($b$) $u_{0,\uparrow}(\vr_\parallel)$ and ($c$) $u_{0,\downarrow}(\vr_\parallel)$. Parameter values are $t=1$, $\mu=0.2$, $\Delta_0=0.1$, and $\xi/a=\hbar v_F /\Delta=10$ on a $L_x\times L_y=400\times 400$ lattice.
}
\label{fig: zero-mode wavefunction}
\end{figure}

The quasi-particle operator
$\hat{\gamma}_{0,k_z}=\int d^2r_\parallel ~ u_{0,k_z\uparrow} \hat{\psi}_{k_z,\uparrow}(\vr_\parallel)
+u_{0,k_z\uparrow} \hat{\psi}_{k_z,\downarrow}(\vr_\parallel)
+u^*_{0,k_z\uparrow} \hat{\psi}^\dagger_{-k_z,\uparrow}(\vr_\parallel)
+u^*_{0,k_z\uparrow}  \hat{\psi}^\dagger_{-k_z,\downarrow}(\vr_\parallel)$
associated with the zero mode solution satisfies $\hat{\gamma}_{0,k_z}=\hat{\gamma}^\dagger_{0,-k_z}$.
Hence, this $k_z$ branch of zero modes is a Majorana zero-energy band. Under open boundary condition along the $z$-direction, we can organize the quasi-particle operators into real Majorana operators as
\begin{align}\label{eq:Majoranasine}
\hat{\gamma}_{k_z}&=\mathcal{N}^{-1}_{k_z}\sum_{z,\sigma}
\int d^2 r_\parallel \sin k_z z \left[u^*_{0, k_z,\sigma}\hat{\psi}_\sigma(\vr)+\text{H.c.}\right],
\nonumber
\end{align}
where $\mathcal{N}_{k_z}$ is the normalization constant. $k_z$ is restricted to $k_S<k_z<k_N$ and takes values $k_z =\frac{n_z \pi}{N_z+1}$, where
$n_z=1,...,N_z$ and $N_z$ is the number of lattice sites in the $z$ direction.

In addition to the above analytic solution which keeps the low-energy electron states, we also solve the vortex problem numerically with the full lattice Hamiltonian by employing Eq. \eqref{eq:Ham} and the onsite pairing gap function $\Delta(\vr_\parallel)$ with the phase winding defined before. The features of the zero-energy vortex state are qualitatively the same. The energy spectrum of the lowest lying energy states at the $q_z=0$ cut is shown in Fig. \ref{fig: zero-mode wavefunction} ($a$), which shows the existence of a zero mode localized at the vortex core. Moreover, the radial wavefunctions $u_{0,q_z=0\uparrow}(\vr)$ and $u_{0,q_z=0\downarrow}(\vr)$ of the zero mode are shown in Figs. \ref{fig: zero-mode wavefunction} ($b$) and ($c$) and are very close to the zeroth and first order Bessel functions, respectively. The phase winding of each component of the zero-energy state is also in agreement with the pattern shown in Eq. \eqref{eq:zero-energysolution}.

{\it Topological Protection of Zero Modes} -- The index theorem \cite{Atiyah1968} is a powerful tool for proving the existence and topological stability of zero modes. As shown by Jackiw and Rebbi, a kink in the spatially dependent mass term of the 1D Dirac equation topologically protects the existence of the zero-energy half-fermion mode \cite{Jackiw1976,Jackiw1981,Cugliandolo1989}. It has also been applied to a 2D topological superconductor to show that a vortex possessing odd-integer winding can host a robust zero-energy Majorana mode \cite{Tewari2007, Tewari2010}.

We perform an analysis for the vortex problem of a monopole superconductor based on the index theorem, as shown in S. M. III. Due to the azimuthal symmetry with respect to the vortex line, the $z$-component of the angular momentum, denoted $m_j$, is a good quantum number, which takes half-integer values due to spin-orbit coupling. As a result of the real space phase winding, Cooper pairing takes place between states with the angular momenta satisfying $m_j+m_j^\prime=1$. Hence, the channel $m_j=m_j^\prime=\frac{1}{2}$ is singled out and decoupled from channels with other angular momenta.  Then, for each momentum cut of $k_z$, there exists a pair of channels with $m_j=m_j^\prime=\frac{1}{2}$ lying on FS$_+$ and FS$_-$, respectively. They form a two-component Nambu spinor and the Hamiltonian can be mapped into a $(1+1)d$ Dirac Hamiltonian
\begin{equation}
\label{eq:monopoleDirac}
\hat{H}=\sum_{k_z>0}\int dx\left[-i\hbar\tilde{v}_F \hat{\phi}^\dagger_{k_z}\tau_z\partial_x\hat{\phi}_{k_z}+ \hat{\phi}^\dagger_{k_z}\tau_y m(x)\hat{\phi}_{k_z}\right],
\end{equation}
where the sum $k_z$ is over all momenta cutting through FS$_+$, and $m(x)$ is a spatially dependent mass term. The details of the derivation can be found in S. M. III, which shows that $m(x)$ is an odd function of spatial coordinate exhibiting a kink configuration. Hence the zero mode at each $k_z$ is topologically protected and belongs to the $m_j=\frac{1}{2}$ channel, consistent with our solution in Eq. \eqref{eq:zero-energysolution}: the spin-up solution has $m_l=0$ and $m_s=\frac{1}{2}$, and the spin-down component has $m_l=1$ and $m_s=-\frac{1}{2}$.

	{\it Majorana Tunnelling.} -- The above analysis can be generalized to multiple well-isolated vortices, with multiple zero modes localized around each vortex. However, in general, Majorana tunnelling between vortices will lift the ground state degeneracy of the ground state \cite{Cheng2009,Cheng2010}. Suppose there are two vortices separated by a distance $R$. In the limit $R\gg k_{F,\parallel}^{-1},\xi$, when $|q_z|<q_F$, inter-vortex tunnelling between Majorana zero modes of the same $k_z$ can lead to an energy shift of
$
    \Delta E\approx  \frac{4\sqrt{2}\hbar v_F |\mathcal{N}_{1,k_z}|^2AB}{\sqrt{\pi}k_{F,\parallel}(1+k_{F,\parallel}^2\xi^2)^{1/4}}\frac{\cos\left(k_{F,\parallel}R+\alpha\right)}{\sqrt{R/\xi}}e^{-R/\xi},
$
where $2\alpha=\arctan k_F\xi$, and the constants $\mathcal{N}_{1,k_z}$, $A$, and $B$ are defined in Eq. \eqref{eq: exact J} of S.M. II. The splitting decays and oscillates with the vortex separation $R$. When $q_F<|q_z|<q_{\text{node}}$, the energy splitting no longer oscillates, and is given by $
    \Delta E \approx -\frac{2\sqrt{2}\hbar v_F|\mathcal{N}_{2,k_z}|^2CD}{\sqrt{\pi}k_{F,\parallel}}\frac{1}{\sqrt{kR}}e^{-k R},
$
where $k=1/\xi-k_{F,\parallel}$, and $\mathcal{N}_{2,k_z}$, $C$, and $D$ are defined in Eq. \eqref{eq: exact I}. When $|q_z|=q_F$, there is no energy splitting due to Majorana tunnelling.

{\it Experimental Realization} -- Along with the rising experimental interest on magnetic Weyl semi-metals \cite{Wang2016,Belopolski2019,Morali2019,Liu2019,Kuroda2017,Yang2017,Higo2018a,Cheng2019,Higo2018b,Nakatsuji2015,Kiyohara2016,Nayak2016,Narita2017,Li2017,Huang2017}, 
we propose to realize monopole superconductivity in a TR-broken Weyl semi-metal, 
for example, 
the antiferromagnet \ce{EuCd_2As2} 
\cite{Wang2019} or magnetic doped Dirac semimetals  
in proximity with an $s$-wave superconductor. 
Under a magnetic field, vortices can form in a monopole superconductor and the zero-energy Majorana modes along the vortex line can be detected by spin-selective Andreev reflection via a scanning tunnelling microscopy experiment 
\cite{Sun2017,Kawakami2015,He2014,Haim2015,Bode2003,Wiesendanger2009,Hu2016,Sun2016}.

{\it Conclusion.} -- We have studied the zero-energy vortex bound states in a monopole superconductor constructed from a Weyl semi-metal with broken TR symmetry in proximity with an $s$-wave superconductor. For every momentum slice labelled by $k_z$ cutting through the FS, there exists a single zero mode. The relation between the real space phase winding of the zero mode and the projected order parameter was discussed. These zero modes are 
topologically protected by the index theorem. 
	
{\it Acknowledgment.} --
C.S. and Y.L. are supported by the U.S. Department of Energy, Office of Basic Energy Sciences, Division of Materials Sciences and Engineering, Grant No. DE-SC0019331. Y.L. also thanks the Institute for Condensed Matter theory of the University of Illinois for hospitality. This work was supported in part by the Alfred P. Sloan Research Fellowships and the Gordon and Betty Moore Foundations EPiQS Initiative through Grant No. GBMF4305.

\bibliographystyle{apsrev4-1}
\bibliography{all}

\balancecolsandclearpage
\section{Supplemental Materials}
  \pagenumbering{arabic}
  \renewcommand{\thepage}{S-\arabic{page}}
\setcounter{equation}{0}
\setcounter{figure}{0}
\setcounter{table}{0}
\setcounter{page}{1}
\makeatletter
\renewcommand{\theequation}{S\arabic{equation}}
\renewcommand{\thefigure}{S\arabic{figure}}
\section{I. Bogoliubov-de Gennes Quasiparticle Spectrum}

In the Nambu basis $\hat{\Psi}_\vk=[\hat{c}_{\vk,\up},\hat{c}_{\vk,\down},\hat{c}^\dagger_{-\vk,\up},\hat{c}^\dagger_{-\vk,\down}]^T$, the Hamiltonian Eq. \eqref{eq:Ham} takes the form $\hat{H}=\frac{1}{2}\sum_{\vk}\hat{\Psi}^\dagger_{\vk}H_{\text{BdG}}(\vk)\hat{\Psi}_{\vk}$, where the one-body Bogoliubov-de Gennes (BdG) Hamiltonian is
	\begin{equation}
	    H_{\text{BdG}}(\vk)=\begin{bmatrix}
	    h(\vk)&\Delta\\
	    \Delta&-h^*(-\vk)
	    \end{bmatrix}.
	\end{equation}
	The quasi-particle energy spectrum is determined by the eigenvalues of $H_{\text{BdG}}(\vk)$,
	\begin{align}
	    E_{\pm,\pm}(\vk)=\pm \bigg[&|h(\vk)|^2+\mu^2+\Delta_0^2\pm 2\sqrt{\mu^2|h(\vk)|^2+\Delta_0^2h_z^2(\vk)}\bigg]^{\frac{1}{2}},
	\end{align}
	where $|h(\vk)|^2=\sum_i h_{i}^2(\vk)$. The quasi-particle spectrum consists of four bands. In the lightly-doped and weak-coupling regime, $|t|\gg |\mu| \gg |\Delta_0|$, there are two linear band crossings at zero-energy in the reduced Brillouin zone ($-\pi\leq k_{x,y}<\pi$,  $0\leq k_z< \pi$) located at $\vec{k}_{N/S}\approx \left(0,0,k_{N/S}\right)$, where $k_N=K_0+ q_{\text{node}}$ and $k_S=K_0-q_{\text{node}}$, $q_{\text{node}}=\sqrt{q_F^2+(\Delta_0/\hbar v_F)^2}$, and the Fermi wavevector $q_F\equiv \mu/(\hbar v_F)$. 

\section{II. Wavefunction of the Zero Mode Without Projection}
	We study the vortex problem in a monopole superconductor and investigate the zero-energy vortex core states. 
    Assuming the vortex line is along the $z$-axis, $k_z$ is conserved because of translation symmetry along the vortex line.
    Define the operator $\hat{\psi}_{k_z\sigma}(\vr_\parallel)=(1/\sqrt{A}) \sum_{k_x,k_y}e^{i(k_x x+k_y y)}\hat{c}_{\vk\sigma}$, where $A$ is the area of the system on the $x$-$y$ plane. For simplicity, we use a low-energy continuum model around $\pm K_0$ for the band Hamiltonian,
\begin{equation}
h_{\pm \mathbf{K}_{0,z}+q_z \hat{\mathbf{z}}}(\vr_\parallel)
=\hbar
v_F(-i\partial_x\sigma_x-i\partial_y\sigma_y\pm q_z\sigma_z)-\mu I,
\end{equation}
where $q_z$ is measured from the band Weyl points
momenta $\pm \mathbf{K_0}$. 
The BdG Hamiltonian reads
\begin{equation}
\hat{H}_{\text{BdG}} = \sum_{k_z>0} \int d^2 r_\parallel
\hat{\Psi}^\dagger_{k_z} (\mathbf{r}_\parallel)
h_{\text{BdG}}(k_z,\mathbf{r}_\parallel)
\hat{\Psi}_{k_z}( \mathbf{r}_\parallel),
\end{equation}
where the four-component Nambu spinor operator is
$\hat{\Psi}^\dagger_{k_z}(\mathbf{r}_\parallel) =
\Big[\hat{\psi}^\dagger_{k_z,\uparrow}(\mathbf{r}_\parallel),
\hat{\psi}^\dagger_{k_z,\downarrow}(\mathbf{r}_\parallel),
\hat{\psi}_{-k_z,\uparrow}(\mathbf{r}_\parallel),
\hat{\psi}_{-k_z,\downarrow}(\mathbf{r}_\parallel) \Big]^T$,
and the summation over $k_z$ only covers $k_z$.
The matrix kernel $h_{\text{BdG}}(k_z,\mathbf{r}_\parallel)$
is defined as
\begin{equation}
h_{\text{BdG}}(k_z,\vr_\parallel)=
\begin{bmatrix}
h_{k_z}(\vr_\parallel)&\Delta(\vr_\parallel)i \sigma_y\\
-\Delta^*(\vr_\parallel)i\sigma_y&-h_{-k_z}^*(\vr_\parallel)
\end{bmatrix}.
\end{equation}
 In order to obtain the quasi-particle excitation spectrum, we perform a Bogoliubov transformation
to solve for the eigenfunction $\psi_{n,k_z}(\mathbf{r}_\parallel)$
\begin{align}
h_{\text{BdG}}(k_z,\mathbf{r}_\parallel)\psi_{n,k_z}(\mathbf{r}_\parallel)
=E_{n,k_z}\psi_{n,k_z}(\mathbf{r}_\parallel),
\end{align}
where $n$ runs over all the eigenstates. Then $\hat{H}_{\text{BdG}}$ is diagonal,
$\hat{H}=\sum_{n,k_z}E_{n,k_z}\hat{\gamma}^\dagger_{n,k_z}
\hat{\gamma}_{n,k_z}$.
	
We seek the zero-energy vortex bound state solutions to the BdG equation. Note that the Hamiltonian possesses the particle-hole symmetry $\mathcal{C}h_{\text{BdG}}(k_z,\mathbf{r}_\parallel)\mathcal{C}^{-1} =-h_{\text{BdG}}^*(-k_z,\mathbf{r}_\parallel)=-h_{\text{BdG}}^*(k_z,\mathbf{r}_\parallel)$,
where $\mathcal{C}=\tau_xK$ and $K$ is the complex conjugation operator. Therefore, for every solution $\psi_{n,k_z}(\mathbf{r}_\parallel)$ to
$H_{\text{BdG}}(k_z,\mathbf{r}_\parallel)$ with energy $E_{n,k_z}$, there exists another solution $\mathcal{C}\psi_{n,k_z}$ with energy $-E_{n,k_z}$. The zero-energy solutions can be arranged to satisfy $\psi_{0,k_z}=\mathcal{C}\psi_{0,k_z}$.
In other words, we look for the zero-energy solutions of the form $\psi_{0,k_z}(\mathbf{r}_\parallel)=[u_{0,k_z\up}(\mathbf{r}_\parallel),u_{0,k_z\down}(\mathbf{r}_\parallel),
u^*_{0,k_z\up}(\mathbf{r}_\parallel),u^*_{0,k_z\down}(\mathbf{r}_\parallel)]^T$.
For every momentum slice at $k_z=+K_0+q_z$, we consider the exponentially decaying solution corresponding to a vortex bound state,
\begin{equation}\label{eq: exact exponential}
\psi_{0,k_z}(\vr_\parallel)=e^{-\frac{1}{\hbar v_F}\int_0^{r_\parallel} d\rho^\prime\Delta(\rho^\prime)}\chi_{k_z}(\vr_\parallel).
\end{equation}
When $|q_z|<q_F$, $\chi_{k_z}(\vr_\parallel)$ is solved analytically as
\begin{equation}\label{eq: exact J}
\chi_{K_0+q_z}(\vr_\parallel)=\mathcal{N}_{1,k_z}
\begin{bmatrix}
A e^{-i\frac{\pi}{4}}J_0(k_{F,\parallel}r_\parallel)\\
B e^{i\frac{\pi}{4}}e^{i\phi_\vr}J_1(k_{F,\parallel}r_\parallel)\\
A e^{i\frac{\pi}{4}}J_0(k_{F,\parallel}r_\parallel)\\
B e^{-i\frac{\pi}{4}}e^{-i\phi_\vr}J_1(k_{F,\parallel}r_\parallel)
\end{bmatrix},
\end{equation}
where $A=\sqrt{1+q_z/q_F}$, $B=\sqrt{1-q_z/q_F}$, $J_{0,1}(z)$ are the zeroth and first order Bessel functions, respectively, and $k_{F,\parallel}=\sqrt{\left|q_F^2-q_z^2\right|}$ is the in-plane component of the Fermi wavevector. 
When $q_F<|q_z|<q_{\text{node}}$, the solutions for $\psi_{0,k_z}$ are non-oscillating, with
    \begin{equation}
    \label{eq: exact I}
    \chi_{K_0+q_z}(\vr_\parallel)=\mathcal{N}_{2,k_z}
    \begin{bmatrix}
    C e^{-i\frac{\pi}{4}}I_0(k_{F,\parallel}r_\parallel)\\
    D e^{i\frac{\pi}{4}}e^{i\phi_\vr}I_1(k_{F,\parallel}r_\parallel)\\
    C e^{i\frac{\pi}{4}}I_0(k_{F,\parallel}r_\parallel)\\
    D e^{-i\frac{\pi}{4}}e^{-i\phi_\vr}I_1(k_{F,\parallel}r_\parallel)
    \end{bmatrix},
    \end{equation}
	where $I_{0,1}(z)$ are respectively the zeroth and first order modified Bessel functions of the first kind, and $C=\sqrt{1+q_z/q_F} \left(\sqrt{-1-q_z/q_F}\right)$ and $D=-\sqrt{-1+q_z/q_F}\left(\sqrt{1-q_z/q_F}\right)$ for $q_z>q_F$ $(q_z<-q_F)$. 
	When $|q_z|>q_{\text{node}}$, there is no zero-energy solution.
	
	\subsection{Normalization Constant}
The states $\psi_{0,k_z}(\vr_\parallel)$ in Eq. \eqref{eq: exact exponential} are normalized according to $\int d^2 \vr_{\parallel} \psi^\dagger_{0,k_z}(\vr_{\parallel})\psi_{0,k_z}(\vr_{\parallel})=1$. When $|q_z|<q_F$, the normalisation constant is
given by
\begin{equation}
    |\mathcal{N}_{1,k_z}|^2=\frac{1}{4\pi}\frac{1}{A^2S_0+B^2S_1},
\end{equation}
where $S_\nu$ is the integral defined by
\begin{align}
S_\nu=\int_0^\infty dr_{\parallel}e^{-\frac{2}{\hbar v_F}\int_0^{r_{\parallel}}dr^\prime \Delta(r^\prime)}J_\nu^2(k_{F,\parallel}r_{\parallel})\approx \int_0^\infty dr_{\parallel} e^{-2r_{\parallel}/\xi}J_\nu^2(k_{F,\parallel}r_{\parallel}),
\end{align}
with $\xi=\hbar v_F/\Delta_0$. The integral can be evaluated to give \cite{Cheng2010,Abramowitz1948}
\begin{align}
S_\nu&=\frac{k_{F,\parallel}^{2\nu}\xi^{2\nu+2}}{4^{2\nu+1}}\frac{\Gamma(2\nu+2)}{[\Gamma(\nu+1)]^2}F\left( \nu+\frac{1}{2},\nu+\frac{3}{2};2\nu+1;-k_{F,\parallel}^2\xi^2\right),
\end{align}
where $\Gamma(z)$ is the gamma function and $F(a,b,c;z)$ the hypergeometric function.

When $q_F<|q_z|<q_{\text{node}}$, the normalization constant is
\begin{equation}
    |\mathcal{N}_{2,k_z}|^2=\frac{1}{4\pi}\frac{1}{C^2T_0+D^2T_1},
\end{equation}
where $T_\nu$ is the integral defined by
\begin{align}
T_\nu=\int_0^\infty dr_{\parallel}e^{-\frac{2}{\hbar v_F}\int_0^{r_{\parallel}}dr^\prime \Delta(r^\prime)}I_\nu^2(k_{F,\parallel}r_{\parallel})\approx \int_0^\infty dr_{\parallel} e^{-2\frac{\Delta_0}{\hbar v_F}r}I_\nu^2(k_{F,\parallel}r_{\parallel}).
\end{align}
Evaluation of the integral gives
\begin{align}
T_\nu&=\frac{k_{F,\parallel}^{2\nu}\xi^{2\nu+2}}{4^{2\nu+1}}\frac{\Gamma(2\nu+2)}{[\Gamma(\nu+1)]^2}F\left( \nu+\frac{1}{2},\nu+\frac{3}{2};2\nu+1;k_{F,\parallel}^2\xi^2\right).
\end{align}
	
\section{III. Mapping to the (1+1)d Dirac Hamiltonian}

In this section we derive the effective $(1+1)d$ Dirac Hamiltonian Eq. \eqref{eq:monopoleDirac} in the main text. We follow the method presented in \cite{Tewari2007,Tewari2010} and generalize it to our case of a monopole superconductor.

\subsection{Weyl Hamiltonian}
The kinetic Hamiltonian is given by the Weyl band structure Hamiltonian
Eq. \eqref{eq:WeylHam}.
Because only low-energy excitations are of interest, the sum over $\vq$
can be restricted to momenta near the two FSs,
\begin{eqnarray}\label{eq: Weyl + and -}
    \hat{H}_{\text{Weyl}}&=&\hat{H}_{\text{Weyl},+}+\hat{H}_{\text{Weyl},-}
    \nonumber\\
    \hat{H}_{\text{Weyl},\pm}&=&\sum_{\vq,\sigma,\sigma^\prime}\hat{c}^\dagger_{\pm \vK_0+\vq,\sigma}h_{\pm,\sigma\sigma^\prime}(\vq)\hat{c}_{\pm \vK_0+\vq,\sigma^\prime},
\end{eqnarray}
where  $h_{\pm}(\vq)=\hbar v_F(q_x\sigma_x+q_y\sigma_y\pm q_z\sigma_z)-\mu I_\sigma$, as defined in the main text. $\hat{H}_{\text{Weyl},\pm}$ are the parts of the band structure Hamiltonian near the nodes $\pm \vK_0$. It is convenient to exploit the cylindrical symmetry of the vortex by utilizing cylindrical coordinates. By keeping $q_z$ discrete and taking the continuum limit in the $q_x$ and $q_y$ directions, the Hamiltonians $\hat{H}_{\text{Weyl},\pm}$ can be recast into
\begin{equation}
\hat{H}_{\text{Weyl},\pm}=
\frac{1}{(2\pi)^2}\sum_{q_z,\sigma,\sigma^\prime}\int_{k_{F,\parallel}(q_z)-\Lambda}^{k_{F,\parallel}(q_z)+\Lambda}
q_\pp d q_\pp \int_0^{2\pi} d\phi_{\vq} \hat{c}^\dagger_{\pm K_0+q_z,\sigma}(q_\pp,\phi_{\vq})h_{\pm,\sigma\sigma^\prime}(q_\pp,\phi_{\vq})\hat{c}_{\pm K_0+q_z,\sigma^\prime}(q_\pp,\phi_{\vq}),
\end{equation}
where $q_\pp=\sqrt{q_x^2+q_y^2}$ and $\phi_{\vq}=\arctan\frac{q_y}{q_x}$ is the azimuthal angle of $\vq$. Here $k_{F,\parallel}(q_z)\equiv\sqrt{q_F^2 -q_z^2}$ is the in-plane Fermi wavevector for the momentum space cut at $q_z$ and $\Lambda$ is a momentum cutoff. The operators $\hat{c}_{k_z,\sigma}(q_\pp,\phi_{\vq})$ satisfy the fermionic anti-commutation relation
\begin{eqnarray}
\Big\{\hat{c}_{k_z,\sigma}(q_\pp,\phi_{\vq}), \hat{c}^\dagger_{k_z^\prime,\sigma^\prime}(q_\pp,\phi_{\vq}) \Big\}
= (2\pi)^2\delta_{k_z,k_z^\prime}\delta_{\sigma,\sigma^\prime}
\frac{1}{q_\pp}\delta(q_\pp-q'_\pp)\delta(\phi_{\vq}-\phi_{{\vq}'}).
\end{eqnarray}

As a result of spin-orbit coupling, the $z$-component of the orbital and spin angular momenta, $m_l$ and $m_s$, are not good quantum numbers, while the total angular momentum in the $z$ direction, $m_j=m_l+m_s$, is. It is therefore convenient to decompose the fermion operator $\hat{c}^\dagger_{k_z,\sigma}(q_\pp,\phi_{\vq})$ in angular momentum channels with definite $m_j$:
\begin{align}
\begin{bmatrix}
\hat{c}_{k_z,\up}(q_\pp,\phi_{\vq})\\
\hat{c}_{k_z,\down} (q_\pp,\phi_{\vq})
\end{bmatrix}
=\frac{1}{\sqrt{q_\pp}}\sum_{m_j}\begin{bmatrix}
e^{i\left(m_j-\frac{1}{2} \right)\phi_{\vq}}\hat{c}_{k_z,\up,m_j}(q_\pp)\\
e^{i\left(m_j+\frac{1}{2} \right)\phi_{\vq}}\hat{c}_{k_z,\down,m_j}(q_\pp)
\end{bmatrix},
\end{align}
where $m_j$ takes on half-integer values and $\hat{c}_{k_z,\sigma,m_j}(q_\pp)$
satisfies the anticommutation relation
\begin{eqnarray}
\acomm{\hat{c}_{k_z,\sigma,m_j}(q_\pp)}
{\hat{c}^\dagger_{k_z^\prime,\sigma^\prime,m_j^\prime}(q'_\pp)}=
2\pi\delta_{k_z,k_z^\prime}\delta_{\sigma,\sigma^\prime}\delta_{m_j,m_j^\prime}
\delta(q_\pp-q'_\pp).
\end{eqnarray}

Performing the partial wave decomposition on $\hat{H}_{\text{Weyl},+}$ and integrating over the azimuthal angle, it becomes
\begin{equation}
\hat{H}_{\text{Weyl},+}=\sum_{q_z}\sum_{m_j}\int \frac{dq_\pp}{2\pi}
\begin{bmatrix}
\hat{c}^\dagger_{K_0+q_z,\up,m_j}(q_\pp)&\hat{c}^\dagger_{K_0+q_z,\down,m_j}(q_\pp)
\end{bmatrix}\begin{bmatrix}
-\mu +\hbar v_F q_z&\hbar v_F q_\pp\\
\hbar v_F q_\pp &-\mu -\hbar v_Fq_z
\end{bmatrix}\begin{bmatrix}
\hat{c}_{K_0+q_z,\up,m_j}(q_\pp)\\
\hat{c}_{K_0+q_z,\down,m_j}(q_\pp)
\end{bmatrix}.
\end{equation}
For fixed $q_z$ and $m_j$, the Hamiltonian can be diagonalized by the unitary transformation
\begin{align}
\begin{bmatrix}
\hat{c}_{K_0+q_z,\up,m_j}(q_\pp)\\
\hat{c}_{K_0+q_z,\down,m_j}(q_\pp)
\end{bmatrix}=
\begin{bmatrix}
\cos\frac{\theta_{\vq}}{2}& -\sin\frac{\theta_{\vq}}{2} \\
\sin\frac{\theta_{\vq}}{2} & \cos\frac{\theta_{\vq}}{2}
\end{bmatrix}
\begin{bmatrix}\label{eq:unitary1}
\hat{f}_{K_0+q_z,-,m_j}(q_\pp)\\
\hat{f}_{K_0+q_z,+,m_j}(q_\pp)
\end{bmatrix},
\end{align}
where $\theta_{\vq}$ is defined via $q_z=q\cos\theta_{\vq}$ and $q_\pp=q\sin\theta_{\vq}$ with $q=\sqrt{q_z^2+q^2_\pp}$. Because the transformation is unitary, the operators $\hat{f}_{K_0+q_z,m_j,\mp}(q_\pp)$ inherit the fermionic anticommutation relations from $\hat{c}_{K_0+q_z,\sigma,m_j}(q_\pp)$. In terms of these operators, the Hamiltonian reads
\begin{equation}
\hat{H}_{\text{Weyl},+}=\sum_{q_z}\sum_{m_j}\int \frac{d q_\pp}{2\pi}
\begin{bmatrix}
\hat{f}^\dagger_{K_0+q_z,-,m_j}(q_\pp)&\hat{f}^\dagger_{K_0+q_z,+,m_j}(q_\pp)
\end{bmatrix}\begin{bmatrix}
-\mu +\hbar v_F q &0\\
0&-\mu -\hbar v_F q
\end{bmatrix}\begin{bmatrix}
\hat{f}_{K_0+q_z,-,m_j}(q_\pp)\\
\hat{f}_{K_0+q_z,+,m_j}(q_\pp)
\end{bmatrix}.
\end{equation}
The Hamiltonian is projected onto the FS by only keeping states near the
Fermi energy, leading to
\begin{equation}\label{eq: Weyl +}
    \hat{H}_{\text{Weyl,+}}=\sum_{q_z}\sum_{m_j}
     \int_{k_{F,\parallel}(q_z)-\Lambda}^{k_{F,\parallel}(q_z)+\Lambda} \frac{d q_\pp}{2\pi} \hat{f}^\dagger_{K_0+q_z,-,m_j}(q_\pp)\left(-\mu+\hbar v_F q \right)\hat{f}_{K_0+q_z,-,m_j}(q_\pp).
\end{equation}


A similar analysis can be performed for $\hat{H}_{\text{Weyl},-}$. After decomposing into angular momentum channels and performing the unitary transformation
\begin{align}\label{eq:unitary2}
\begin{bmatrix}
\hat{c}_{-K_0-q_z,\up,m_j}(q_\pp)\\
\hat{c}_{-K_0-q_z,\down,m_j}(q_\pp)
\end{bmatrix}=
\begin{bmatrix}
-\sin\frac{\theta_{\vq}}{2} &\cos\frac{\theta_{\vq}}{2} \\
\cos\frac{\theta_{\vq}}{2}& \sin\frac{\theta_{\vq}}{2}
\end{bmatrix}
\begin{bmatrix}
\hat{f}_{-K_0-q_z,-,m_j}(q_\pp)\\
\hat{f}_{-K_0-q_z,+,m_j}(q_\pp)
\end{bmatrix},
\end{align}
the Hamiltonian becomes
\begin{equation}
\hat{H}_{\text{Weyl},-}=\sum_{q_z}\sum_{m_j}\int \frac{d q_\pp}{2\pi}
\begin{bmatrix}
\hat{f}^\dagger_{-K_0-q_z,-,m_j}(q_\pp)&\hat{f}^\dagger_{-K_0-q_z,+,m_j}(q_\pp)
\end{bmatrix}\begin{bmatrix}
-\mu -\hbar v_F q &0\\
0&-\mu +\hbar v_F q
\end{bmatrix}\begin{bmatrix}
\hat{f}_{-K_0-q_z,-,m_j}(q_\pp)\\
\hat{f}_{-K_0-q_z,+,m_j}(q_\pp)
\end{bmatrix}.
\end{equation}
Projecting onto the FS, the Hamiltonian reads
\begin{equation}\label{eq: Weyl -}
    \hat{H}_{\text{Weyl},-}=\sum_{q_z}\sum_{m_j}
    \int_{k_{F,\parallel}(q_z)-\Lambda}^{k_{F,\parallel}(q_z)+\Lambda} \frac{d q_\pp}{2\pi}
    \hat{f}^\dagger_{-K_0-q_z,+,m_j}(q_\pp)\left(-\mu+\hbar v_F q \right)\hat{f}_{-K_0-q_z,+,m_j}(q_\pp).
\end{equation}


\subsection{Pairing Hamiltonian}
The mean-field pairing Hamiltonian with a vortex of winding number $+1$ imposed is 
\begin{equation}
\hat{H}_{\Delta}=\sum_{k_z>0,\sigma,\sigma^\prime} \int d^2r_\parallel
~\Delta(\vr_\parallel)e^{i\phi_\vr} \hat{\psi}^\dagger_{k_z,\sigma}(\vec{r}_\parallel)[i\sigma_y]_{\sigma\sigma^\prime}\hat{\psi}^\dagger_{-k_z,\sigma^\prime}(\vec{r}_\parallel)+\text{H.c.},
\end{equation}
where $\Delta(r_\parallel)=\Delta_0\tanh(r_\parallel/\xi)$. Again the sum over $k_z$ only covers $k_z>0$.


Switching to the momentum representation, $\hat{\psi}^\dagger_{k_z,\sigma}(\vec{r}_\parallel)= \int \frac{d^2\vq_{\pp}}{(2\pi)^2} e^{-i\vec{q}_\pp\cdot\vec{r}_\parallel}\hat{c}^\dagger_{k_z,\sigma}(\vec{q}_\pp)$, the pairing Hamiltonian becomes
\begin{align}
\hat{H}_{\Delta}&=\sum_{k_z>0,\sigma,\sigma^\prime}\int \frac{d^2\vq_\pp}{(2\pi)^2} \frac{d^2\vq^\prime_\pp}{(2\pi)^2}
\underbrace{\int d^2r\Delta(\rho_\vr)e^{i\phi_\vr}
e^{-i(\vec{q}_\pp+\vec{q}_\pp^\prime)\cdot\vec{r}_\parallel}}_{I(\vec{q}_\pp+\vec{q}^\prime_\pp)}
\hat{c}^\dagger_{k_z,\sigma}(q_\pp,\phi_\vq)[i\sigma_y]_{\sigma\sigma^\prime}
\hat{c}^\dagger_{- k_z,\sigma^\prime}(q_\pp^\prime,\phi_{\vq'})+ \text{H.c.}\nonumber.
\end{align}
The integral $I(\vq_\pp+\vq'_\pp)$ has been evaluated in Refs. \cite{Tewari2007,Tewari2010}, which we repeat here for completeness. Writing the integral in cylindrical coordinates as
\begin{equation}
I(\vq_\pp+\vq'_\pp)=\int d^2r_\parallel \Delta(r_\parallel)e^{i\phi_{\vr}}
e^{-i |\vq_\pp+\vq'_\pp| r_\parallel\cos\left(\phi_{\vr}-\phi_{\vq+\vq^\prime}\right)}
\end{equation}
and making the change of variables $\phi_\vr\to \phi_\vr+\phi_{\vq+\vqp}+\frac{\pi}{2}$, it becomes
\begin{align}
I(\vq_\pp+\vq'_\pp)&=e^{i\left(\phi_{\vq+\vqp}+\frac{\pi}{2} \right)}
\int r_\parallel dr_\parallel \Delta(r_\parallel)
\int d\phi_\vr e^{i|\vq_\pp+\vq_\pp^\prime|r_\parallel\sin\left( \phi_\vr+\phi_\vr\right)}.
\end{align}
Using the integral representation for Bessel functions of integer order $J_n(z)=\frac{1}{2\pi}\int_{-\pi}^\pi d\tau e^{i(z\sin \tau-n\tau)}$, the $\phi_\vr$ integral can be written as $2\pi J_{-1}(|\vq_\pp+\vq^\prime_\pp|r_\parallel)$. Finally, performing the integral over $r_\parallel$, we obtain
\begin{equation}
I(\vq_\pp+\vq'_\pp)=-2\pi i\Delta_0 \frac{q_{\pp}e^{i\phi_\vq}+q_{\pp}^\prime e^{i\phi_{\vqp}}} {|\vq_\pp+\vq'_\pp|^3}.
\end{equation}
\

Substituting in the integral, the pairing Hamiltonian becomes
\begin{align}
\hat{H}_{\Delta}=-2\pi i\Delta_0 \sum_{k_z>0}\int \frac{d^2 \vq_\pp}{(2\pi)^2}
\frac{d^2 \vq^\prime_\pp}{(2\pi)^2} \frac{q_{\pp}e^{i\phi_\vq}+q_{\pp}'e^{i\phi_{\vqp}}}{|\vq_\pp+\vq'_\pp|^3}
[\hat{c}^\dagger_{k_z,\up}(\vq_\pp)\hat{c}^\dagger_{- k_z,\down}(\vq'_\pp)-(\up\leftrightarrow\down)]+\text{H.c.}.
\end{align}
Because $|\vq_\pp+\vq_\pp^\prime|=\sqrt{q_\pp^2+q_\pp^{'2}-2q_\pp q'_\pp \cos(\phi_{\vq}-\phi_{\vqp})}$ is periodic in $\phi_{\vq}-\phi_{\vqp}$, we can expand it in angular momentum channels:
\begin{align*}
\frac{1}{|\vq_\pp+\vq_\pp^\prime|^3}=
\sum_{n}u_n(q_\pp,q_\pp^\prime)e^{in(\phi_\vq-\phi_{\vqp})},
\end{align*}
where $n$ takes on integer values. Since $|\vq_\pp+\vq_\pp^\prime|$ is invariant under swapping $q_\pp$ and $q_\pp^\prime$, the Fourier coefficients satisfy $u_n(q_\pp,q_\pp^\prime)=u_{n}(q_\pp^\prime,q_\pp)$. Furthermore, because it only depends on $\cos(\phi_\vq-\phi_{\vqp})$, which is even in $\phi_\vq-\phi_{\vqp}$, we have $u_{n}(q_\pp,q_\pp^\prime)=u_{-n}(q_\pp,q_{\pp}^\prime)$ and real Fourier coefficients. Expanding the fermion operators in angular momentum channels as well and performing the angular integrals, the pairing Hamiltonian
becomes
\begin{align}
\hat{H}_{\Delta}&=-2\pi i \Delta_0
\sum_{k_z}\sum_n\int \frac{d q_\pp}{2\pi} \frac{d q'_\pp}{2\pi}
\sqrt{q_\pp q'_\pp} u_{n}(q_\pp,q'_\pp)  \bigg[ q_\pp
 \hat{c}^\dagger_{k_z,\up,n+\frac{3}{2}}(q_\pp)
 \hat{c}^\dagger_{-k_z,\down,-n-\frac{1}{2}}(q'_\pp)
 +q'_\pp \hat{c}^\dagger_{k_z,\up,n+\frac{1}{2}}(q_\pp)
 \hat{c}^\dagger_{-k_z,\down,-n+\frac{1}{2}}(q'_\pp)
 \nonumber \\
 &-q_\parallel \hat{c}^\dagger_{k_z,\down,n+\frac{1}{2}}(q_\pp)\hat{c}^\dagger_{-k_z,\up,-n+\frac{1}{2}}(q_\pp^\prime)-q_\pp ^\prime \hat{c}^\dagger_{k_z,\down,n-\frac{1}{2}}(q_\p)\hat{c}^\dagger_{-k_z,\down,-n+\frac{3}{2}}(q_\pp^\prime)\bigg]+\text{H.c.}.
 \label{eq:pairing}
\end{align}
The $m_j=\frac{1}{2}$ channel ($n=-1$ in first term, $n=0$ in second and third, and $n=1$ in fourth)  in Eq. \eqref{eq:pairing} pairs with itself and is decoupled from the rest. The Hamiltonian for this angular momentum channel reads
\begin{equation}
\hat{H}_{\Delta,m_j=\frac{1}{2}}
=-i\sum_{k_z} \int \frac{d q_\pp}{2\pi} \frac{dq^\prime_\pp}{2\pi}
\left[h_\Delta(q_\pp,q_\pp^\prime)\hat{c}^\dagger_{k_z,\up,\frac{1}{2}}(q_\pp)\hat{c}^\dagger_{-k_z,\down,\frac{1}{2}}(q_\pp^\prime) -h_{\Delta}(q_\pp^\prime,q_\pp)\hat{c}^\dagger_{k_z,\down,\frac{1}{2}}(q_\pp)\hat{c}^\dagger_{-k_z,\up,\frac{1}{2}}(q_\pp^\prime)\right]+\text{H.c.},
\end{equation}
where
\begin{equation}
    h_{\Delta}(q_\pp,q_\pp^\prime)=2\pi  \Delta_0 \sqrt{q_\pp q_\pp^\prime} \left[q_\pp u_{-1}(q_\pp,q_\pp^\prime)+q_\pp^\prime u_0(q_\pp,q_\pp^\prime) \right].
\end{equation}
In what follows we deal exclusively with the $m_j=\frac{1}{2}$ angular momentum channel so the $m_j$ index will henceforth be neglected.

Since only the low-energy excitations are of interest, we project the pairing Hamiltonian onto the FSs. Using the transformations in Eqs. \eqref{eq:unitary1} and \eqref{eq:unitary2}, we write $\hat{c}^\dagger_{k_z,\up}(q_\pp)\approx\cos\frac{\theta_\vq}{2}\hat{f}^\dagger_{k_z,-}(q_\pp)$,  $\hat{c}^\dagger_{k_z,\down}(q_\pp)\approx\sin\frac{\theta_\vq}{2}\hat{f}^\dagger_{k_z,-}(q_\pp)$, $\hat{c}^\dagger_{-k_z,\up}(q_\pp^\prime)\approx \cos\frac{\theta_{\vqp}}{2}\hat{f}^\dagger_{-k_z,+}(q_\pp^\prime)$, and $\hat{c}^\dagger_{-k_z,\down}(q_\pp^\prime)\approx \sin\frac{\theta_{\vqp}}{2}\hat{f}^\dagger_{-k_z,+}(q_\pp^\prime)$. Making these substitutions, the Hamiltonian becomes
\begin{equation}\label{eq: pair1}
\hat{H}_{\Delta}=-i\sum_{k_z}\int\frac{dq_\pp}{2\pi}\frac{dq^\prime_\pp}{2\pi}A(q_\pp-q_\pp^\prime)\hat{f}^\dagger_{k_z,-}(q_\pp)\hat{f}^\dagger_{-k_z,+}(q_\pp^\prime)+\text{H.c.},
\end{equation}
where
\begin{equation}
   A(q_\pp-q_\pp^\prime)=h_{\Delta}(q_\pp,q_{\pp}^\prime)\cos\frac{\theta_\vq}{2}\sin\frac{\theta_{\vq^\prime}}{2}-h_{\Delta}(q_\pp^\prime,q_\pp) \sin\frac{\theta_\vq}{2}\cos\frac{\theta_{\vq^\prime}}{2} 
\end{equation}
is a real function that is antisymmetric under the exchange of $q_\pp$ and $q_\pp^\prime$.

\subsection{Fourier Transform}
Finally, combining the above results, we can obtain an expression for the full Hamiltonian $\hat{H}=\hat{H}_{\text{Weyl}}+\hat{H}_{\Delta}$.  Using Eqs. \eqref{eq: Weyl + and -}, \eqref{eq: Weyl +}, \eqref{eq: Weyl -}, and \eqref{eq: pair1}, $\hat{H}$ takes the form
\begin{align}
   \hat{H}&=\sum_{q_z}\int_{-\Lambda}^{\Lambda} \frac{dq_\pp}{2\pi} 
   \begin{bmatrix}
   \hat{f}^\dagger_{k_z,-}(q_\pp )&\hat{f}_{-k_z,+}(q_\pp)
   \end{bmatrix}\begin{bmatrix}
  \hbar v_F (q-q_F) 
   &0 \\
   0&
   -\hbar v_F (q-q_F)
   \end{bmatrix}
   \begin{bmatrix}
   \hat{f}_{k_z,-}(q_\pp)\\
   \hat{f}^\dagger_{-k_z,+}(q_\pp)
   \end{bmatrix}\nonumber\\
   &+\sum_{q_z}\int_{-\Lambda}^{\Lambda} \frac{dq_\pp}{2\pi} 
   \int_{-\Lambda}^{\Lambda}
   \frac{dq_{\pp}^\prime}{2\pi} 
   \begin{bmatrix}
   \hat{f}^\dagger_{k_z,-}(q_\pp )&\hat{f}_{-k_z,+}(q_\pp)
   \end{bmatrix}\begin{bmatrix}
  0
   &-iA(q_\pp -q'_\pp) \\
   iA(q_\pp -q'_\pp)&
  0
   \end{bmatrix}
   \begin{bmatrix}
   \hat{f}_{k_z,-}(q'_\pp)\\
   \hat{f}^\dagger_{-k_z,+}(q'_\pp)
   \end{bmatrix}.
\end{align}
We expand $\hbar v_F (q-q_F)=\hbar \tilde{v}_F \delta q_\pp$, 
where $\tilde{v}_F = v_F\sin\theta_{\vq}$ and $\delta q_\pp=q_\pp -k_{F,\parallel}(q_z)$, i.e.,
$\delta q_\pp$ is set to zero at the Fermi radius at a fixed $q_z$.
Then we have
\begin{align}
   \hat{H}&=\sum_{q_z}\int_{-\Lambda}^\Lambda \frac{d \delta q_\pp}{2\pi} 
   \begin{bmatrix}
   \hat{f}^\dagger_{k_z,-}(\delta q_\pp )&\hat{f}_{-k_z,+}( \delta q_\pp)
   \end{bmatrix}\begin{bmatrix}
  \hbar \tilde{v}_F\delta q_\pp
   &0 \\
  0&
   - \hbar \tilde{v}_F\delta q_\pp
   \end{bmatrix}
   \begin{bmatrix}
   \hat{f}_{k_z,-}( \delta q_\pp)\\
   \hat{f}^\dagger_{-k_z,+}(\delta q_\pp)
   \end{bmatrix}\nonumber\\
   &+\sum_{q_z}\int_{-\Lambda}^\Lambda \frac{d \delta q_\pp}{2\pi} 
   \int_{-\Lambda}^\Lambda \frac{d\delta q'_\pp}{2\pi} 
   \begin{bmatrix}
   \hat{f}^\dagger_{k_z,-}(\delta q_\pp )&\hat{f}_{-k_z,+}( \delta q_\pp)
   \end{bmatrix}\begin{bmatrix}
 0
   &-iA(\delta q_\pp - \delta q'_\pp) \\
   iA(\delta q_\pp - \delta q'_\pp)&
   0
   \end{bmatrix}
   \begin{bmatrix}
   \hat{f}_{k_z,-}( \delta q'_\pp)\\
   \hat{f}^\dagger_{-k_z,+}(\delta q'_\pp)
   \end{bmatrix}.
\end{align}
Since we are interested in the low-energy physics, we send the momentum cutoff $\Lambda\to \infty$.

Define the Fourier transformations
\begin{align}
\hat{f}_{k_z,-}(\delta q_\pp)&=\int_{-\infty}^\infty dx e^{-i \delta q_\pp x} \hat{f}_{k_z,-}(x),&
\hat{f}_{-k_z,+}(\delta q_\pp)&=\int_{-\infty}^\infty dx e^{-i \delta q_\pp x} \hat{f}_{-k_z,+}(x).
\end{align}
In this new representation, the Hamiltonian is
\begin{equation}
    \hat{H}=\sum_{q_z}\int dx\left[-i\hbar \tilde{v}_F
    \hat{\phi}^\dagger_{k_z}\tau_z\partial_x\hat{\phi}_{q_z}+ \hat{\phi}^\dagger_{k_z}\tau_y m(x)\hat{\phi}_{k_z}\right],
\end{equation}
where $\hat{\phi}_{k_z}(x)=\left[\hat{f}_{k_z,-}(x),
\hat{f}^\dagger_{-k_z,+}(-x) \right]^T$, and the mass function
\begin{equation}
    m(x)\equiv\int \frac{d \delta q_\pp}{2\pi} A( \delta q_\pp )e^{i\delta q_\pp x}
\end{equation}
is an odd function in $x$ because $A(\delta q_\pp)$ is an odd function
in $\delta q_\pp$.
This is the derivation of Eq. \eqref{eq:monopoleDirac} in the main text.


\end{document}